# SS 433: C II emission from the disk photosphere

M. G. Bowler

Department of Physics, University of Oxford, Keble Road, Oxford OX1 3RH
e-mail:michael.bowler@physics.ox.ac.uk

**Abstract**
The Galactic microquasar SS 433 is a member of a binary system but there is a lack of data on the orbital velocities of the components. The emission lines of the C II doublet at 7231 and 7236 Å have been tracked nightly over two orbital cycles. The spectra are adequate to establish that these lines are eclipsed by the companion and hence to extract a measure of the orbital velocity of the compact object; the lines are formed in the disk photosphere. This velocity is 176 ± 13 km s$^{-1}$. Could XSHOOTER do better?

**Introduction**
The optical spectrum of the microquasar SS 433 has been assiduously searched for emission and absorption lines exhibiting Doppler shifts varying in time with a period of 13 days. Most lines exhibit some 13 day periodicity (which is how the binary nature of the system was first revealed, Crampton, Cowley and Hutchings 1980) but in the vast majority of cases the phase of the oscillation does not match the variation in projected velocity of either the compact object or the donor (Crampton & Hutchings 1981; Gies et al 2002). Absorption lines have been reported with correct phasing for formation close to the companion, but there is no consistency among the orbital velocities inferred for the companion. For example, Hillwig & Gies (2008) report 58 km s$^{-1}$ and Cherepashchuk et al (2005) 132 km s$^{-1}$.

The notable emission line He II at 4486 Å has revealed a source orbiting with the phase of the compact object (Crampton & Hutchings 1981). This source is totally eclipsed by the companion at orbital phase 0 and has been assigned to the base of the jets of SS 433; the periodic Doppler shifts yield an orbital velocity for the compact object about the binary centre of mass of 176 ± 13 km s$^{-1}$ (Fabrika 2004; see also Fabrika 1997; Fabrika & Bychkova 1990). It is on the basis of these observations that a value of 7.7 $M_\odot$ has been assigned to the mass function of SS 433. The He II emission is not uncomplicated; the signal consists of a (comparatively) narrow component (FWHM ~ 1000 km s$^{-1}$) assigned to a gas stream within the system in addition to the broader component attributed to the base of the jets, which dominates when the disk is most turned toward us. The Doppler amplitude for lumped He II data is only about 120 km s$^{-1}$ (D'Odorico et al 1991; see Fabrika 1997, 2004 for detailed discussions). Because of these complications additional data on the orbital speed of the compact object are scarcely redundant; data on the circumbinary disk yield a system mass of around 40 $M_\odot$ for an orbital velocity of 175 km s$^{-1}$ and imply that the compact object is a rather massive stellar black hole (Blundell, Bowler & Schmidtobreick 2008).
In a major study of the optical spectrum of SS 433 Gies et al (2002) noted that a weak emission feature matching the C II doublet at 7231,7236 Å not only exhibited a periodic Doppler shift but that shift was maximally to the red when the compact object recedes fastest (orbital phase 0.75). Their extracted orbital speed is 162 ± 29 km s$^{-1}$, entirely consistent with the He II results. However, their data are sparse and drawn from a number of observing periods.

A paper concerned with extracting properties of the system from X-ray observations (Cherepashchuk at al 2013) emphasised the need for better data on He II and the C II doublet; this is as true today as it was then. The only new determinations that I know of are to be found in Kubota et al (2010) and in Picchi & Shore et al (2020). The latter extracts the speed of the compact object from He II, cobbling together 4 data points from Kubota et al (2010) and 7 from XSHOOTER data, two of which were discarded because of concerns about the complexity of the line. The data points fitted cover orbital phase 0.08 to 0.87. They quote a result 160± 3 km s$^{-1}$. I am doubtful that such a small fitted error can be a reasonable estimate of the uncertainty.

No new data on C II emission have been published, but Picchi et al (2020) stimulated me to resurrect an analysis of 15 years ago and prepare this note.

**La Silla observations of C II**

Between August and late September 2004 were accumulated spectra of SS 433 on a nightly basis, over the range 5800 Å to 8700 Å using the ESO 3.6 m telescope on La Silla. These observations were primarily directed to the moving jets (Blundell, Bowler & Schmidtobreick 2007) but contain an almost unbroken sequence of nights, covering two orbital periods, during which the C II doublet at 7231, 7236 Å was clearly present in every spectrum and not contaminated by the moving lines from the jets. (Fig.1 of Blundell et al 2007 sets this doublet in the context of the full optical spectrum). Although this feature is weak and occurs in a region of the spectrum which suffers from telluric absorption lines, it is clear from the raw data that the position of the doublet oscillates with a period of 13 days and with phase appropriate to a source moving with the compact object. It is also clear that the intensity relative to the neighbouring He I emission at 7281 Å drops during eclipse of the compact object and its disk by the donor. Examples of the far red part of the spectrum, containing C II and the neighbouring He I line, are shown in Fig.1. The lowest panel is for JD 2453000 + 253.5, where the orbital phase is 0.891, the compact object moving away from us at rather less than maximum recessional velocity. The uppermost panel is for JD +258.5, orbital phase 0.275, when the C II doublet is almost maximally blue shifted. The C II doublet has moved to the blue by 315 km s$^{-1}$ during this 5 day period; in contrast the He I line (which is formed in the circumbinary disc and the large scale wind) has scarcely shifted.

 I defined the centre of the C II doublet to be half way between the intersection of the outside edges of the doublet and the smooth curve drawn through the continuum (the horizontal line within each panel of Fig.1). An alternative definition, measuring half way across a width of 10 Å, made little difference. No more elaborate treatment is justified, the weak doublet being dragged back and forth across a noisy continuum broken up by the sky lines, and the error on individual measurements cannot be determined reliably. From the spread in values where two or more measurements are close in orbital phase (together with different methods of determining the centre of the doublet) I estimate errors of ~ 30 km s$^{-1}$ and in some instances considerably more. For the purposes of fitting I have adopted an error of 40 km s$^{-1}$.

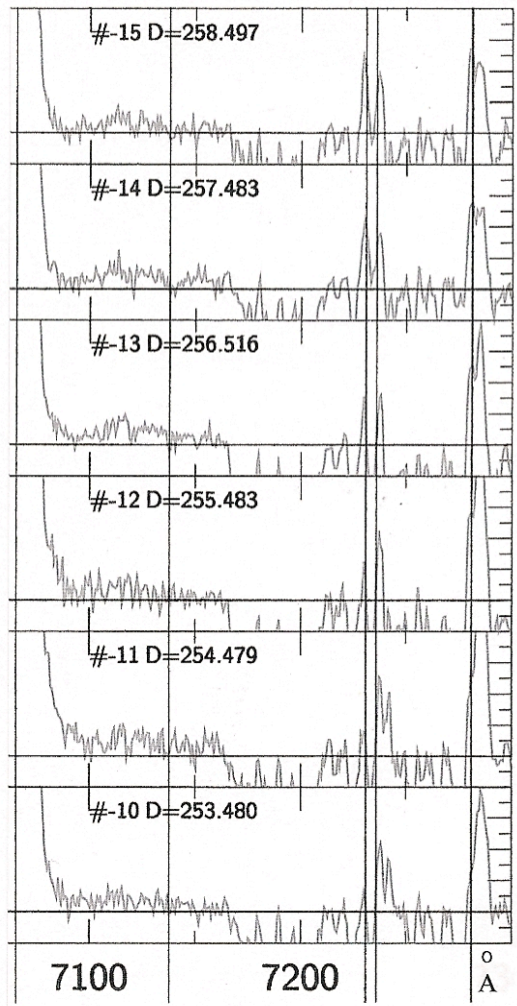

Fig. 1. Examples of 6 consecutive spectra containing both the C II doublet and He I 7281 Å. The lowest panel is for JD +253.5 (just after reddest) and the uppermost almost half a period later at JD 258.5 (bluest). The C II doublet shifts to the blue by 315 km s$^{-1}$ over this period; the He I line (which is split because of contributions from both the red and blue sides of the circumbinary disc) here shifts by about 30 km s$^{-1}$. The double line marks the rest wavelengths of the C II doublet at 7231.32 and 7236.42 Å. On the left of the figure is the tail of the much stronger He I line at 7065 Å.

Fig.2 shows the Doppler shift of the centre of the C II doublet, determined as described above, over two orbits. These data show an oscillation of period approximately 13 days and with maximum and minimum redshifts matching well orbital phases 0.75 (compact object receding) and 0.25 (compact object approaching). The fitted curve is a sinusoidal oscillation of period 13.08 days.

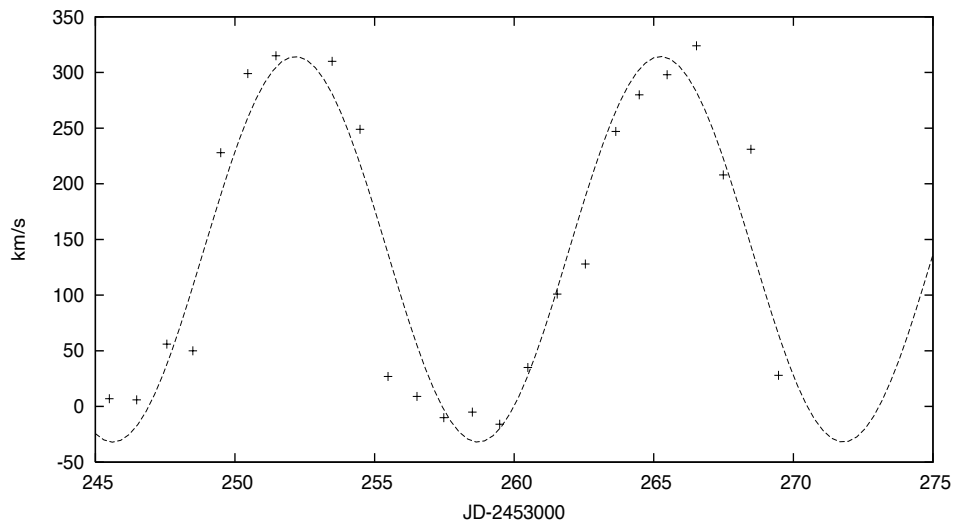

Fig. 2. The redshift of the C II emission doublet is displayed as a function of time over two orbits. Orbital phase zero, when the companion eclipses the compact object, occurs very close to JD + 255 and + 268. The curve is the best fit to a sinusoidal oscillation of period 13.08 days and has an amplitude of 173 km s$^{-1}$.

The orbital velocity, the mean and the phase were variables in the fit. The fitted amplitude is 173 ± 13 km s$^{-1}$ and the mean red shift 141 km s$^{-1}$, with a statistical uncertainty of ∼ 9 km s$^{-1}$. According to the ephemeris of Goranskii et al (1998) orbital phase 0, corresponding to maximum eclipse by the companion, occurs at JD + 254.90 and + 267.98. The fitted curve shown in Fig.2 is shifted later by 0.53 ± 0.15 days. [If the Goranskii phase is used as a constraint, the amplitude becomes 167 ± 16 km s$^{-1}$.]

He I emission lines are formed in wind from the accretion disk and more particularly in the circumbinary disk. The intensities of the red and blue rims of the latter oscillate with orbital phase (Blundell, Bowler & Schmidtobreick 2008) and hence the shape leans alternately to the red or the blue - this can be clearly seen for the He I lines at 6678 Å and 7065 Å in Fig.1 of Schmidtobreick & Blundell (2006). The result is that the contribution from the circumbinary disc shows some oscillation which is not due to Doppler shift. Indeed, when the He I 7281 Å line is measured in the same way as the C II doublet an oscillation with an amplitude of about 30 km s$^{-1}$ is found, about a mean red shift of ∼ 110 km s$^{-1}$.

Being formed in the more remote regions of the SS 433 system, He I emission lines are not eclipsed. C II emission is eclipsed by the donor companion at orbital phase zero.

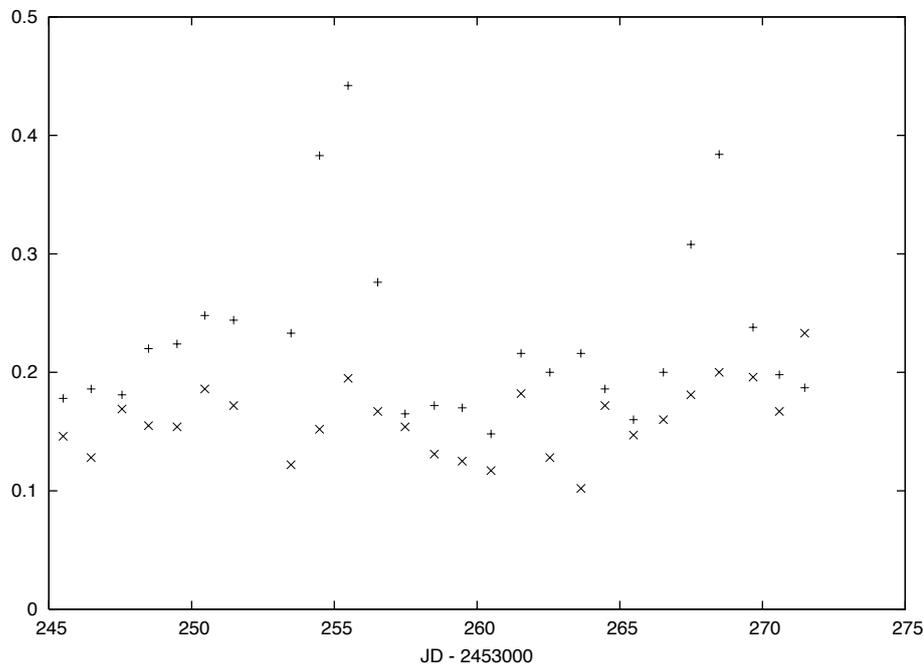

Fig. 3. The ratio of the height of C II emission to the height of the continuum is shown by x and the ratio of He I 7281 Å to the continuum by + . The latter has doubled on JD +255.5 and 268.5 when the continuum is eclipsed by the companion. The figure shows that the source of C II emission is eclipsed to the same extent as the continuum source and hence that C II emission is formed in the photosphere of the accretion disk.

Fig.3 shows that the source of C II emission is eclipsed to the same extent as the source of the continuum. These data do not have absolute photometry and so I employed He I emission (primarily at 7281 Å) to normalise. For the C II doublet, the mean height of the two components above the local continuum was divided by the height of the continuum. For comparison, the height of the He I 7281 Å line above its local continuum was divided by the continuum height and these two ratios are plotted as a function of time in Fig.3. The ratio of He I to continuum increased by approximately a factor of two during eclipse at JD 2453000 + 255.5 and + 268.5 because the continuum source is eclipsed and the He I source is not. The ratio of C II to the continuum does not change. Thus the C II source matches closely the distribution in space of the continuum source and is eclipsed by the companion to much the same extent.

 C II Conclusions

The C II doublet is emitted from a source with fairly small velocity dispersion; the full width at half height (insofar as it can be estimated) is ∼ 10 Å , close to the separation of the two components plus the FWHM resolution of 2.2 Å . The source shows Doppler shifts with phase characteristic of motion with the compact object and above all the source is eclipsed by the companion like the source of the continuum radiation. The source of C II must be in the photosphere of the accretion disk of SS 433 and consequently the Doppler amplitude should yield a true measure of the orbital velocity of the compact object about the binary centre of mass. After correction for orbital inclination, that orbital velocity is 176 km s$^{-1}$ with a notional accuracy of 13 km s$^{-1}$. This is in excellent agreement with the He II data and the orbital velocity of the compact object about the centre of mass of the binary can be taken as well established.

The same cannot be said for the recessional velocity of the SS 433 system as a whole. The data from He II have a mean redshift of approximately zero; 27 ±13 km s$^{-1}$ from Crampton & Hutchings (1981) and -22 ± 14 km s$^{-1}$ from Fabrika & Bychkova (1990). The C II emission line would seem a good candidate for determining the true systemic velocity, but Gies et al (2002) report 200 ± 20 km s$^{-1}$, to be compared with the La Silla result of 141 ± 9 km s$^{-1}$. C II results are also higher than implied by the properties of the circumbinary disc. The He I line at 7281 Å has a mean redshift of ∼ 110 km s$^{-1}$ and this is likely to be skewed high because the red edge of the circumbinary disc is usually brighter than the blue (Schmidtobreick & Blundell 2006). In Balmer Hα the mean redshift of the circumbinary disc is ∼ 70 km s$^{-1}$ (Blundell, Bowler & Schmidtobreick 2008) and again this may be to the red of the true systemic velocity. The mean redshift of the C II doublet seems curiously large.

**A note on Si II emission lines**

A figure in Picchi, Shore et al (2020) displays a spectrum in the region of the Si II line at 6347.1 Å. Their primary interest was an absorption trough (apparently consistent with the companion), but the emission line is stronger. I wondered if this was also from the region of the compact object and therefore attempted an analysis similar to that on C II. The La Silla Si II emission lines are weaker than C II and it was not possible to obtain a result of much accuracy. This emission line has the phase of the compact object and the orbital velocity is ∼ 175 km s$^{-1}$ but I estimate the uncertainty as ∼ 40 km s$^{-1}$. This is not much use, but it raises the question of whether XSHOOTER data could do better with the two emission doublets, C II and Si II.

**Acknowledgement.** The data used in this note, in the form exemplified by Fig.1, are from the remarkable Blundell/Schmidtobreick set. They were made available to me about 15 years ago. Figs. 2 & 3 of this note illustrate once more the power of this sequence of daily spectra.